\documentclass[a4paper,11pt]{article}
\usepackage{pos}

\usepackage{url}
\usepackage{fancybox}
\usepackage[english]{babel}
\usepackage{times}
\usepackage{graphicx}
\usepackage{psfrag}
\usepackage{subfigure}

\usepackage{tabulary}
\usepackage{comment}

\usepackage{comment}
\definecolor{darkred}{rgb}{0.4,0.0,0.0}
\definecolor{darkbred}{rgb}{0.7,0.0,0.0}
\definecolor{darkgreen}{rgb}{0.0,0.4,0.0}
\definecolor{darkblue}{rgb}{0.0,0.0,0.4}
\definecolor{darkmagenta}{rgb}{0.55, 0.0, 0.55}
\definecolor{bole}{rgb}{0.47, 0.27, 0.23}
\definecolor{LightBlue}{rgb}{0.5,0.5,1}

\usepackage{braket}
\usepackage[T1]{fontenc}
\usepackage{amsmath}
\usepackage{amsthm}
\usepackage{url}
\usepackage[utf8]{inputenc}
\usepackage{dcolumn}% eqnarray table columns on decimal point
\usepackage{bm}% bold math
\usepackage{url} 
\usepackage{grffile}
\usepackage{xcolor}
\usepackage{booktabs}
\usepackage{verbatim}

\usepackage{listings}
\newcommand{\ben}{ \begin{enumerate}}
\newcommand{\een}{\end{enumerate} }
\newcommand{\bbl}{\begin{block}{} \begin{center}}
\newcommand{\ebl}{\end{center} \end{block}}
\newcommand{\bref}{\begin{flushright} \begin{tiny}}
\newcommand{\eref}{\end{tiny} \end{flushright}}

\usepackage{amsmath}
\usepackage{amssymb}
\usepackage{braket}
\usepackage{array,multirow}
\usepackage{tikz}
\usepackage{dsfont}
\usetikzlibrary{snakes}
\usetikzlibrary{math}
\usetikzlibrary{calc}

\tikzset{decoration={snake,amplitude=0.5mm,segment length=2mm,
		post length=0mm,pre length=0mm}}
\newcommand{\midarrow}{\tikz \draw[-stealth] (0,0) -- +(0,.1);}
\tikzset{every path/.style={line width=0.02 cm}}

\newcommand{\BBdisconnected}[5] % {scale}{xoffset}{yoffset}{quark flavour 1}{quark flavour 2}
{
	\begin{tikzpicture}[baseline={([yshift=-.7ex]current bounding box.center)}]
		\draw[color=#4] (0,#1*1.3) -- node {\midarrow}(0,0);
		\draw[decorate, color=#5] (0,#1*1.3) to [bend left=45] (0,0);
		\draw[color=#4] (#1,1.3*#1) -- node {\midarrow} (#1,0);
		\draw[decorate, color=#5] (#1,0) to [bend left=45] (#1,#1*1.3);
		\node at (0.3*#1,1.2*#1) {#2};
		\node at (0.65*#1,1.2*#1) {#3};
	\end{tikzpicture}
}

\newcommand{\BBconnected}[5] % {scale}{xoffset}{yoffset}{quark flavour 1}{quark flavour 2}
{
	\begin{tikzpicture}[baseline={([yshift=-.65ex]current bounding box.center)}]
		\draw[color=#4] (0,#1*1.3) -- node {\midarrow} (0,0);
		\draw[decorate, color=#5] (0,#1*1.3) to (#1,0);
		\draw[color=#4] (#1,1.3*#1) -- node {\midarrow} (#1,0);
		\draw[decorate, color=#5] (0,0) to  (#1,#1*1.3);
		\node at (0.3*#1,1.2*#1) {#2};
		\node at (0.65*#1,1.2*#1) {#3};
	\end{tikzpicture}
}

\newcommand{\Bmeson}[5] % {scale}{xoffset}{yoffset}{quark flavour 1}{quark flavour 2}
{
	\begin{tikzpicture}[baseline={([yshift=-.65ex]current bounding box.center)}]
		\draw[color=#4] (0,0) -- node {\midarrow}(0,#1*1.3);
		\draw[decorate, color=#5] (0,#1*1.3) to [bend left=45] (0,0);
		\node at (0.3*#1,1.2*#1) {#2};
	\end{tikzpicture}
}

\usepackage{graphicx} % Required for including images
\usepackage[font=small,labelfont=bf, figurename=Fig.]{caption} 
\usepackage{amsmath} % For typesetting math
\usepackage{amssymb}
\usepackage{lipsum}  % For dummy text
\usepackage{color, colortbl}
\selectcolormodel{HTML}
\definecolor{CeFEMA}{rgb}{0.0, 0.6, 1.0}
\definecolor{grey}{RGB}{0.06, 0.05, 0.03}
\definecolor{DarkRed}{RGB}{100,0,10}
\definecolor{OIST}{HTML}{C80019} 
\definecolor{OISTL}{HTML}{F9A8AF} 
\definecolor{concColor}{HTML}{4E95BA} 
\definecolor{BGcolour}{RGB}{40,40,40}
\usepackage{palatino} % Use the Palatino font

\newcommand{\Tr}{\text{Tr}}
\graphicspath{
	{operators/},{figures/}, {posterPlots/}} % Directory in which figures are stored

\title{%
	The spectrum of open confining strings in the large-$N_c$ limit
}
\author*[a]{Alireza Sharifian}
\author[b]{Andreas Athenodorou}
\author[a]{Pedro Bicudo}

\emailAdd{ alireza.sharifian@tecnico.ulisboa.pt}
\emailAdd{ a.athenodorou@cyi.ac.cy}
\emailAdd{ bicudo@tecnico.ulisboa.pt}

\FullConference{The 41st International Symposium on Lattice Field Theory (LATTICE2024)\\
	28 July - 3 August 2024\\
	Liverpool, UK\\}

\affiliation[a]{CeFEMA and Physics department, Instituto Superior Técnico, Av.\ Rovisco Pais, 1049 Lisboa, Portugal}
\affiliation[b]{Computation-based Science and Technology Research Center, The Cyprus Institute, Cyprus}

\abstract{
We compute the spectra of open flux tubes formed between a static quark-antiquark pair for various 
gauge groups in the large-$N_c$ limit, focusing on different symmetries manifested by the quantum numbers of angular momentum, parity, and charge conjugation. Specifically, we present spectra from $N_c=3$ up to $N_c=6$ and for eight different irreducible representations of the symmetry characterizing the flux tube. In this study, we employed an anisotropic Wilson action, a large number of suitable spacial operators, and solved the generalized eigenvalue problem (GEVP) to obtain a significant number of excitations for different combinations of flux tube quantum numbers. The spectra are compared with the Nambu-Goto string model, revealing novel phenomena such as the presence of massive axions in the flux tube spectrum. We find that the mass of the axion in the open flux tube spectra sector is consistent with the mass obtained in the closed flux tube sector.
}

\begin{document}
%%%%%%%%%%%%%%%%%%%%%%%%%%%%%%%%%%%%%%%%%%%%%%%%%%%%%%%%%%%%%%%%%%%%%%%%%%%%%

\maketitle

%SSSSSSSSSSSSSSSSSSSSSSSSSSSSSSSSSSSSSSSSSSSSSSSSSSSSSSSSSSSSSSSSSSSSSSSSSSSSSSSSSSSSSSSSSSSSSSSSSSSSSSSSSSSSSS
%SSSSSSSSSSSSSSSSSSSSSSSSSSSSSSSSSSSSSSSSSSSSSSSSSSSSSSSSSSSSSSSSSSSSSSSSSSSSSSSSSSSSSSSSSSSSSSSSSSSSSSSSSSSSSS
%SSSSSSSSSSSSSSSSSSSSSSSSSSSSSSSSSSSSSSSSSSSSSSSSSSSSSSSSSSSSSSSSSSSSSSSSSSSSSSSSSSSSSSSSSSSSSSSSSSSSSSSSSSSSSS
\section{Preamble}
In confining gauge theories such as \( SU(N_c) \) gluodynamics, the energy of a gauge field from color charges can be compressed into flux tubes, forming confining strings. Understanding the dynamics of the worldsheet of these strings is crucial for comprehending color confinement. Over the past decade, significant advancements have been made in understanding the effective string theoretical description of the {\it closed} flux tube~\cite{Dubovsky:2013gi,Athenodorou:2021vkw}. Notably, the dynamics of the worldsheet theory have been derived from lattice data in a model-independent manner. This analysis revealed the presence of several string-like states and a massive excitation known as the worldsheet axion. However, a similar investigation is lacking for the case of the {\it open} flux tube. In this work, we shed light on the large-\( N_c \) limit of the spectrum of open flux-tubes and the existence of axions in their worldsheet. 

\section{The Nambu-Goto string}
\label{sec:Nambu_Goto}
A long flux tube can be understood as a low-energy bosonic string. It describes the dynamics of translational $(D-2)$ Goldstone bosons, which exhibit a non-linearly realized $ISO(1,D-1)$ Poincaré symmetry. This symmetry is spontaneously broken to the $ISO(1,1) \times O(D-2)$ subgroup. Consequently, the physics of this string is expected to be described by an action most naturally expressed in terms of $D$ worldsheet scalar fields $X^{\mu}$, which represent the embedding of the string worldsheet into the target spacetime. The leading term of this action, in terms of the string tension $\sigma$, is the simplest bosonic string model, namely the Nambu-Goto action given by:
\begin{align}
S=-\sigma \int \mathrm{d}^2x \sqrt{ -\det h_{\alpha\beta}}\, ,
\end{align}
where $\sigma$ is the string tension and $h_{\alpha\beta}$ is the induced metric of the world-sheet, also known as the first fundamental form $h_{\alpha\beta} = \partial_{\alpha}X^{\mu} \partial_{\beta} X_{\mu}$.

%The simplest string model is the Nambu-Goto model, defined by the action,
%\begin{align}
%S=-\sigma \int \mathrm{d}^2\Sigma, 
%\end{align}
%where $\sigma$ is the string tension and $\Sigma$ is the surface of the worldsheet swept by the string. T
The energy spectrum of an open relativistic string with length $R$ and fixed ends is obtained by quantizing Nambu-Goto is given by: 
%	\begin{align}
%	&V(R)=\sqrt{\sigma^2 R^2+2\pi\sigma (N-(D-2)/24)},
%	\label{eq:nambu_goto}
%	\end{align}
    \begin{align}  
	&V(R)=\sqrt{\sigma^2 R^2+2\pi\sigma \left(N-\frac{(D-2)}{24} \right)},
	\label{eq:nambu_goto}
	\end{align}
where $N$ is the quantum number for string vibrations and $D$ is the dimension of space time. This expression is known as the Arvis potential \cite{ARVIS1983106}.

%SSSSSSSSSSSSSSSSSSSSSSSSSSSSSSSSSSSSSSSSSSSSSSSSSSSSSSSSSSSSSSSSSSSSSSSSSSSSSSSSSSSSSSSSSSSSSSSSSSSSSSSSSSSSSS
%SSSSSSSSSSSSSSSSSSSSSSSSSSSSSSSSSSSSSSSSSSSSSSSSSSSSSSSSSSSSSSSSSSSSSSSSSSSSSSSSSSSSSSSSSSSSSSSSSSSSSSSSSSSSSS
%SSSSSSSSSSSSSSSSSSSSSSSSSSSSSSSSSSSSSSSSSSSSSSSSSSSSSSSSSSSSSSSSSSSSSSSSSSSSSSSSSSSSSSSSSSSSSSSSSSSSSSSSSSSSSS
\section{The Axionic String Ansatz (ASA)}
\label{subsec:ASA}

In the case of closed flux tubes, the universal predictions of the Thermodynamic Bethe Ansatz (TBA) have been shown to accurately describe a large sector of confining string excitations. This has been confirmed with high precision in both $D=4$ and $D=3$ dimensions \cite{Athenodorou:2010cs,Athenodorou:2011rx,Athenodorou:2018sab,Athenodorou:2022pmz}. However, at shorter string lengths $R$, most states exhibit increasing deviations from these predictions, indicating the influence of non-universal corrections. TBA analysis has allowed for the determination of the first non-trivial Wilson coefficient for $D=3$ confining strings~\cite{Dubovsky:2014fma,Chen:2018keo}. Interestingly, some $D=4$ states show significant deviations from universal TBA predictions even at relatively large $R$, suggesting the presence of an additional massive excitation on the string worldsheet, commonly referred to as the "string axion" \cite{Dubovsky:2013gi,Dubovsky:2014fma}.

This suggests the existence of an additional term in the worldsheet action that governs the leading-order interactions between the worldsheet axion and the transverse Goldstone modes. The corresponding action is given by:
\begin{equation}
    S_\phi= \int d^2 \sigma \sqrt{-h} \left( -\frac{1}{2} (\partial \phi)^2 - \frac{1}{2} m^2 \phi^2 + \frac{Q_{\phi}}{4} h^{\alpha \beta} \epsilon_{\mu \nu \lambda \rho} \partial_\alpha t^{\mu \nu} \partial_\beta t^{\lambda \rho} \phi \right) \,,
    \label{axion_interaction}
\end{equation}
where \(\phi\) represents the pseudoscalar worldsheet axion, and \(t^{\mu \nu} = \frac{\epsilon^{\alpha \beta}}{\sqrt{-h}} \partial_\alpha X^\mu \partial_\beta X^\nu\). The axion mass \(m\) can be determined from Monte Carlo simulations of the 4D \(SU(N_c)\) Yang-Mills confining flux tube spectrum using TBA analysis, yielding values of \(m=1.82(2)\) for $SU(3)$, \(m=1.65(2)\) for $SU(5)$ and  \(m=1.65(2)\) for $SU(6)$ \cite{Athenodorou:2024loq}.
%SSSSSSSSSSSSSSSSSSSSSSSSSSSSSSSSSSSSSSSSSSSSSSSSSSSSSSSSSSSSSSSSSSSSSSSSSSSSSSSSSSSSSSSSSSSSSSSSSSSSSSSSSSSSSS
%SSSSSSSSSSSSSSSSSSSSSSSSSSSSSSSSSSSSSSSSSSSSSSSSSSSSSSSSSSSSSSSSSSSSSSSSSSSSSSSSSSSSSSSSSSSSSSSSSSSSSSSSSSSSSS
%SSSSSSSSSSSSSSSSSSSSSSSSSSSSSSSSSSSSSSSSSSSSSSSSSSSSSSSSSSSSSSSSSSSSSSSSSSSSSSSSSSSSSSSSSSSSSSSSSSSSSSSSSSSSSS
\section{The quantum numbers of the flux tube}
\label{sec:quantum_numbers}
\begin{figure}[!h]
   \begin{minipage}{0.48\textwidth}
     \centering
              \includegraphics[scale=1.4]{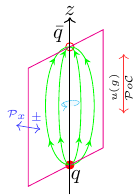}
             \caption{	The possible quantum numbers are:
	$\Sigma_g^+$, $\Sigma_g^-$, $\Sigma_u^+$,$\Sigma_u^-$,$\Pi_g$,$\Pi_u$, $\Delta_u$,$\Delta_g$, $\ldots$
\label{fig_qnumbers}}
    \end{minipage}\hfill
   \begin{minipage}{0.48\textwidth}
     \centering
	     \includegraphics[scale=1.33]{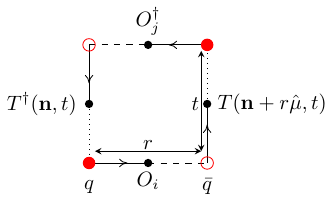}
	     \caption{A closed loop corresponds to the entry $\mathcal{C}_{i,j}(r,t)$ of the Wilson correlation matrix.\label{fig:wilson_loop}}
   \end{minipage}
\end{figure}

There are three constants of motion whose eigenvalues are used to label the quantum state of the flux tube, illustrated in Fig. \ref{fig_qnumbers}:
		\begin{itemize}
			\item $z$-component of angular momentum $\Lambda=0,1,2,3,\ldots$, they are typically denoted by greek letters  $\Sigma,\ \Pi,\ \Delta,\ \Phi, \ldots$, respectively
			\item Combination of the charge conjugation and spatial inversion with respect to the mid point of the charge axis operators  $\mathcal{P}o\mathcal{C}$, its  eigenvalues  $\eta=1,-1$, they are denoted by $g,\ u$, respectively.
			\item For $\Sigma (\Lambda=0)$, reflection with respect to a plan containing the charge axis, $\mathcal{P}_x$, with  eigenvalues $\epsilon=+,-$.
		\end{itemize}

\section{Our lattice framework to study the flux tube}
\label{sec:quantum_numbers}

In this work, we focus on the large-distance behavior of the excited spectrum, deliberately avoiding short-distance effects. To achieve this, we employ the Wilson action discretized on anisotropic lattices, which is expressed as:
\begin{equation}
	S_\text{Wilson} = \beta\left(\frac{1}{\xi} \sum_{x,s>s'} W_{s,s'} + \xi \sum_{x,s} W_{s,t} \right), \label{eq:wilsonAniso}
\end{equation}
where $\beta = 6/g^2$ is the inverse coupling constant and $\xi$ represents the bare anisotropy factor, defined as the ratio of the spatial to temporal lattice spacings $(a_s/a_t)$. Furthermore, with $s$, $s'$ we denote spatial links in different directions. Here, $P_{s,s'}$ represents the spatial plaquettes, and $P_{s,t}$ the spatial-temporal plaquettes.

By using an anisotropic action, we achieve finer lattice spacing in the temporal direction while maintaining a coarser spacing in  spatial directions. This results in more data points for the effective mass plot and enables the study of larger distances. The ensembles generated with this action are listed in Table \ref{tab:ensemble}. We ensure that simulations are ergodic by investigating the topological charge and its evolution. 

\begin{table}[t!]
	\centering
	\begin{tabular}{c||ccccccccc}
		\hline \hline
		Ensemble&$N_c$&$\beta$&$\xi$&$\xi_r$&Volume&$a_t\sqrt{\sigma}$&$a_s\sqrt{\sigma}$
		\\
		\hline \hline
		$W_{3,2}$&$3$&$5.90$&$2.00$&$2.1737(4)$&$24^3\times 48$&$0.1403(8)$&$0.3049(17)$
		\\
			$W_{3,4}$&$3$&$5.70$&$4.00$&$4.5363(8)$&$24^3\times 96$&$0.0887(4)$&$0.4023(16)$
		\\
			$W_{4,2}$&$4$&$10.70$&$2.00$&$2.0958(2)$&$24^3\times 48$&$0.1680(2)$&$0.3521(5)$
		\\
			$W_{4,4}$&$4$&$10.40$&$4.00$&$4.2940(5)$&$24^3\times 96$&$0.1018(3)$&$0.4372(12)$
		\\
			$W_{5,2}$&$5$&$17.20$&$2.00$&$2.0596(1)$&$24^3\times 48$&$0.1451(3)$&$0.2988(5)$
		\\
			$W_{5,4}$&$5$&$16.50$&$4.00$&$4.1853(3)$&$24^3\times 96$&$0.1025(0)$&$0.4288(1)$
		\\
			$W_{6,4}$&$6$&$24.00$&$4.00$&$4.1274(2)$&$24^3\times 96$&$0.0996(0)$&$0.4113(2)$
		\\
		\hline \hline
	\end{tabular}
 \vspace{0.5cm}
\caption{Details for the ensembles generated using the anisotropic Wilson action. $\xi_r$ is the normalized anisotropic factor, and each ensemble includes 1000 configurations. All ensembles have been smeared 100 times using multihit in the temporal direction and APE smearing ($\alpha=0.3, n_s=20$) in  spatial directions. \label{tab:ensemble}}
\end{table}

To compute the spectra of the flux tube, we first calculate the Wilson correlation matrix $\mathcal{C}(r, t)$. The entry $\mathcal{C}_{i,j}(r, t)$ of the Wilson correlation matrix is the expectation value of spatial-temporal closed loops (Fig. \ref{fig:wilson_loop}), where the spatial sides are replaced with operators $O_i$ and $O_j$ that have identical symmetry to the flux tube of interest.

\begin{figure}[!t]
   \begin{minipage}{0.48\textwidth}
     \centering
	\includegraphics[scale=0.9]{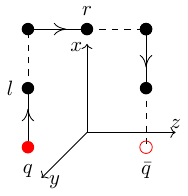}
\caption{Staples for $\Sigma_g^+$,  $\Pi_u$, and $\Delta_g$ flux tubes.\label{fig:op_oi}}
    \end{minipage}\hfill
   \begin{minipage}{0.48\textwidth}
     \centering
	\includegraphics[scale=0.9]{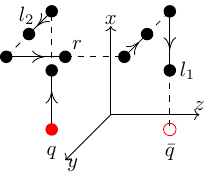}
\caption{Staples for $\Sigma_g^-$ flux tubes.}
   \end{minipage}
\end{figure}

\begin{figure}[!t]
   \begin{minipage}{0.48\textwidth}
     \centering
	\includegraphics[scale=0.9]{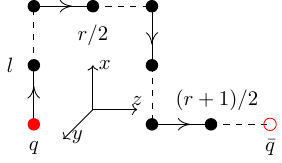}
\caption{Staples for $\Sigma_u^+$,$O^{\Pi_g}$,and  $O^{\Delta_u}$   flux tubes.}
    \end{minipage}\hfill
   \begin{minipage}{0.48\textwidth}
     \centering
	\includegraphics[scale=0.9]{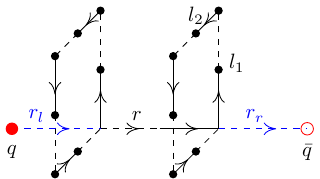}
\caption{Staples for $\Sigma_u^-$ flux tubes. \label{fig:op_loop}}   \end{minipage}
\end{figure}

For example, $O_i$ and $O_j$ can be constructed using the staples shown in Figs. \ref{fig:op_oi} to \ref{fig:op_loop}.

Next, we find the generalized eigenvalues $\lambda$ of the Wilson correlation matrix:
\begin{align}
	\mathcal{C}(r, t)\vec{\nu}_n = \lambda_n(r, t) \mathcal{C}(r, t_0) \vec{\nu}_n,
\end{align}
where we set $t_0 = 0$. Consequently, we obtain a set of time-dependent eigenvalues $\lambda_n(t)$ for each $r$. We then order the eigenvalues and plot the effective mass defined as:
\begin{align}
	E_i(r) = \ln \frac{\lambda_i(r, t)}{\lambda_i(r, t+1)}.
\end{align}
The value of the plateau in the effective mass plot corresponds to the energy $E_i(r)$ of the flux tube.

\begin{comment}
\section{Topological Charge}
The topological charge is defined as
\begin{eqnarray}
	Q = \frac{1}{32\pi^2} \int d^4 x \: \epsilon_{\mu\nu\rho\sigma} \Tr\left[G_{\mu\nu}(x)G_{\rho\sigma}(x)\right] \,.
	\label{eq:Q_continuum_def}
\end{eqnarray}
where $G_{\mu\nu}$ is the field strength tensor of the gauge field. In simulations of large-$N_c$ gauge fields, it is crucial to track the topological charge to ensure the ergodicity of the Monte Carlo process. As $N_c$ increases, the topological charge tends to freeze. In Fig.~\ref{fig:tc_oscillation}, we show the time progression of the topological charge for the largest $N_c=6$, confirming that in our simulation, the topological charge is not frozen, and thus the simulation is ergodic.
\end{comment}

In this work, we use the operators, already used in Ref.~\cite{Sharifian:2023} to study $SU(3)$ spectra. Figs. \ref{fig:op_oi} to \ref{fig:op_loop} are the shapes of the stables used to build up the operators for different flux tube symmetries.

%SSSSSSSSSSSSSSSSSSSSSSSSSSSSSSSSSSSSSSSSSSSSSSSSSSSSSSSSSSSSSSSSSSSSSSSSSSSSSSSSSSSSSSSSSSSSSSSSSSSSSSSSSSSSSS
%SSSSSSSSSSSSSSSSSSSSSSSSSSSSSSSSSSSSSSSSSSSSSSSSSSSSSSSSSSSSSSSSSSSSSSSSSSSSSSSSSSSSSSSSSSSSSSSSSSSSSSSSSSSSSS
%SSSSSSSSSSSSSSSSSSSSSSSSSSSSSSSSSSSSSSSSSSSSSSSSSSSSSSSSSSSSSSSSSSSSSSSSSSSSSSSSSSSSSSSSSSSSSSSSSSSSSSSSSSSSSS
\section{Results on the spectrum of the open flux-tube}

\begin{figure}[t!]
	\includegraphics[scale=0.7500]{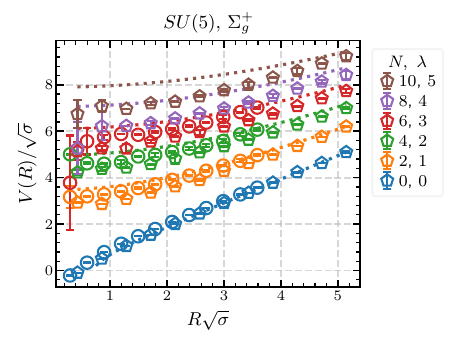}
	\includegraphics[scale=0.7500]{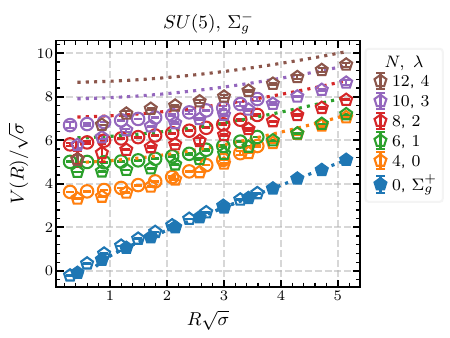}
	\includegraphics[scale=0.7500]{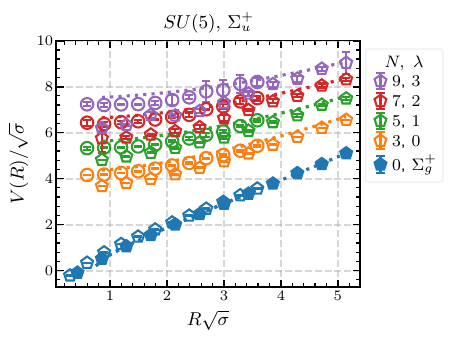}
	\includegraphics[scale=0.7500]{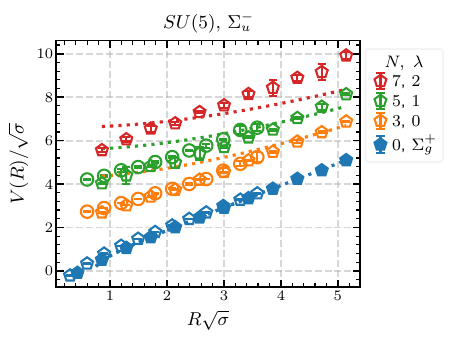}
	\includegraphics[scale=0.7500]{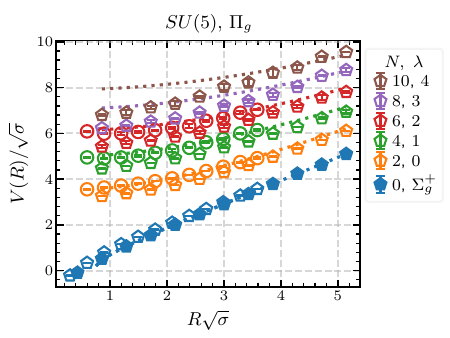}
	\includegraphics[scale=0.7500]{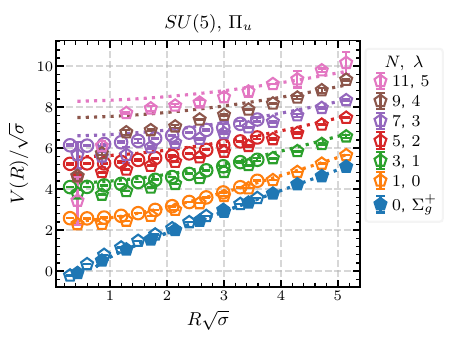}
	\includegraphics[scale=0.7500]{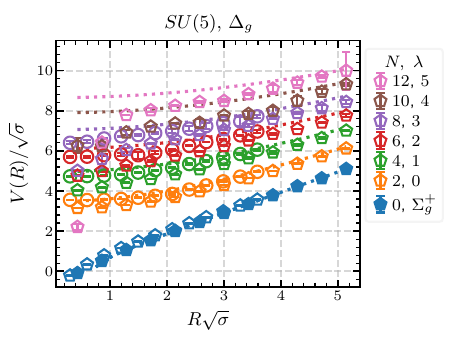} \hspace{95pt}
	\includegraphics[scale=0.7500]{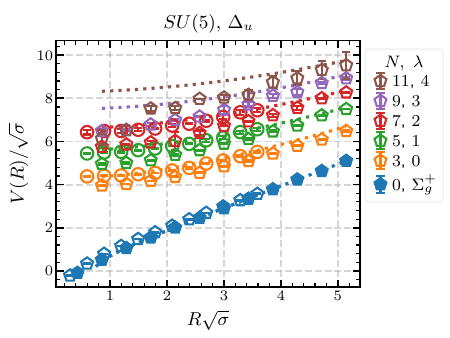}
\caption{Eight SU(5) spectra.  Polygonal shapes in each figure show the simulation with $\xi=4$, while circle markers indicate $\xi=2$. $N$ is the quantum number as defined in Eq. \ref{eq:nambu_goto}, and $\lambda$ is the excitation number.   \label{fig:SU5spectra}}
\vspace{-0.5cm}
\end{figure}

\begin{figure}[t!]
	\includegraphics[scale=0.7500]{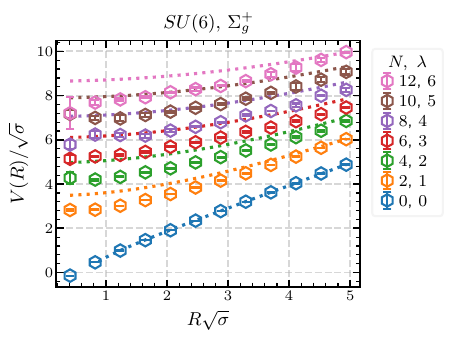}
	\includegraphics[scale=0.7500]{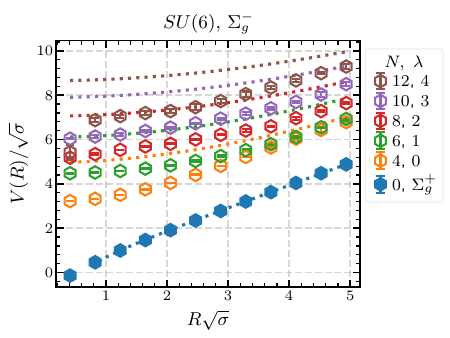}
	\includegraphics[scale=0.7500]{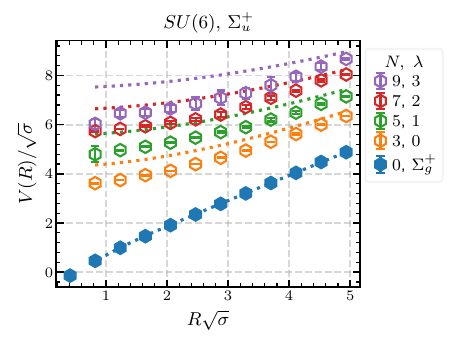}
	\includegraphics[scale=0.7500]{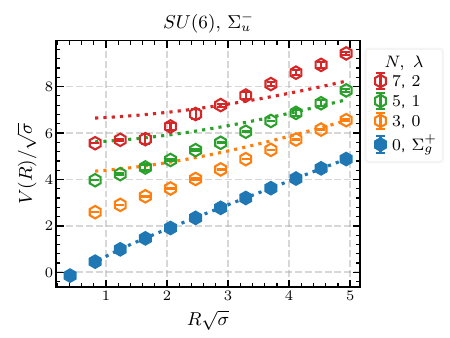}
	\includegraphics[scale=0.7500]{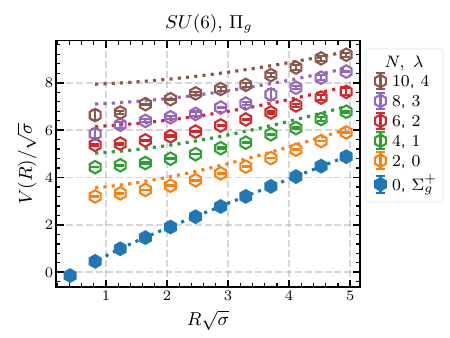}
	\includegraphics[scale=0.7500]{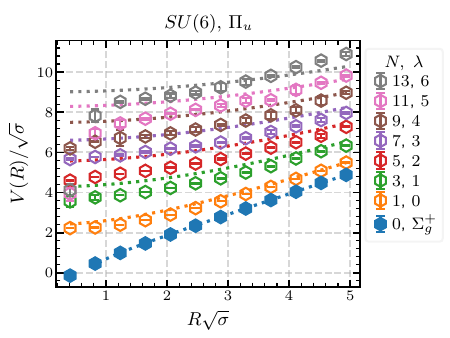}
	\includegraphics[scale=0.7500]{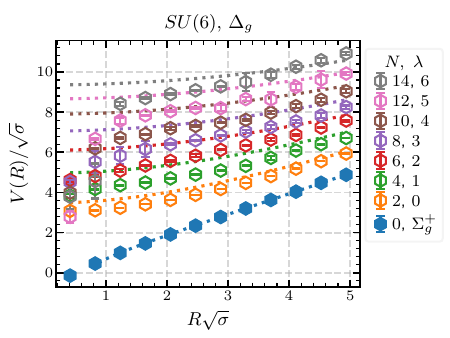}  \hspace{95pt}
	\includegraphics[scale=0.7500]{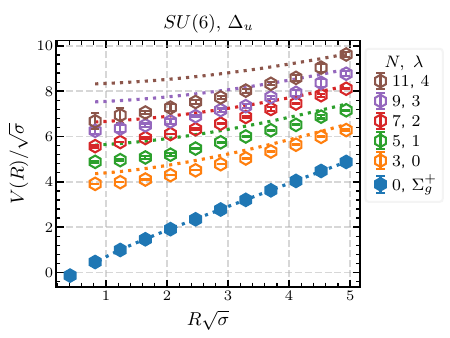}
\caption{ Eight SU(6) spectra. Polygonal shapes in each figure show the simulation with $\xi=4$, while circle markers indicate $\xi=2$. $N$ is the quantum number as defined in Eq. \ref{eq:nambu_goto}, and $\lambda$ is the excitation number.   \label{fig:SU6spectra}}
\vspace{-0.65cm}
\end{figure}

In this section, we present our findings on the spectra for $N_c = 5$ and $N_c = 6$, the largest values of $N_c$ considered in this study, across the eight irreducible representations: $\Sigma^+_g$, $\Sigma^-_g$, $\Sigma^+_u$, $\Sigma^-_u$, $\Pi_g$, $\Pi_u$, $\Delta_g$, and $\Delta_u$. The spectra for these representations are displayed in Figures \ref{fig:SU5spectra} and \ref{fig:SU6spectra} for $N_c = 5$ and $N_c = 6$, respectively. For $SU(5)$, we present results at two different lattice spacings to examine whether the spectrum’s behavior changes as we approach the continuum limit.

The absolute ground state corresponds to the $\Sigma^+_g$ representation, closely approximated by the Arvis potential to a remarkable degree. Deviations of the leading order in $1/(R\sqrt{\sigma})^4$, as derived in \cite{Aharony:2009gg}, are expected. However, the focus of this work is not on subleading corrections in each state's energy but rather on comparing the spectra with the Nambu-Goto string to investigate potential axion signatures. This ground state can be interpreted as the vacuum state $|0\rangle$, with no phonon operators acting on it, implying that no scattering processes occur along the string’s worldsheet. Thus, we expect the spectrum to be well-represented by the Nambu-Goto model.

The first higher excited state one can find in the spectrum is ground state of $\Pi_u$ which is presented in orange colour in the third row from top and second column. This state agrees to a quite good accuracy with the Nambu-Goto state with $N=1$ meaning that it "carries" only one phonon with one unit of momentum. This is not surprising if one considers again the contributions coming from scattering processes along the world-sheet do not exist. Using the creation/annihilation phonon operators, this state has the form $\alpha^{\dagger}_{\pm}|0\rangle$ ~\cite{Sharifian:2023}.

\begin{figure}[t!]
\vspace{-0.5cm}
	\includegraphics[scale=0.7500]{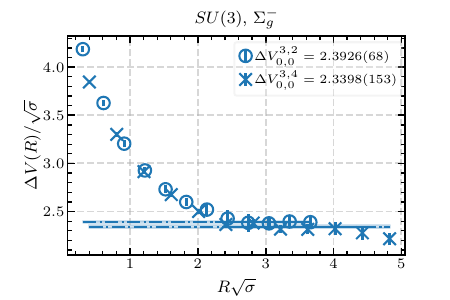}
	\includegraphics[scale=0.7500]{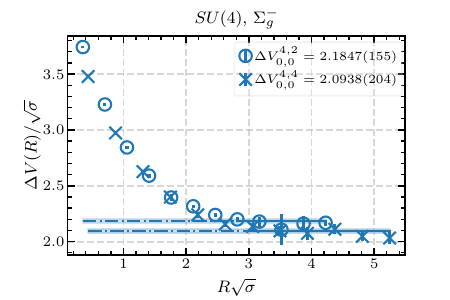}
	\includegraphics[scale=0.7500]{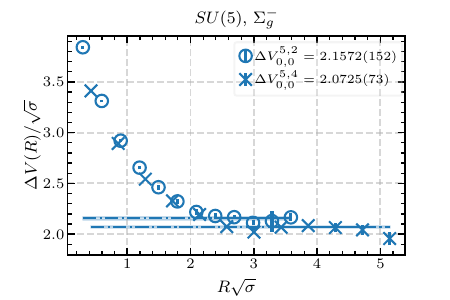}
	\includegraphics[scale=0.7500]{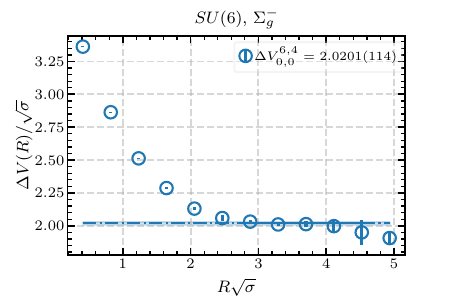}
	\includegraphics[scale=0.7500]{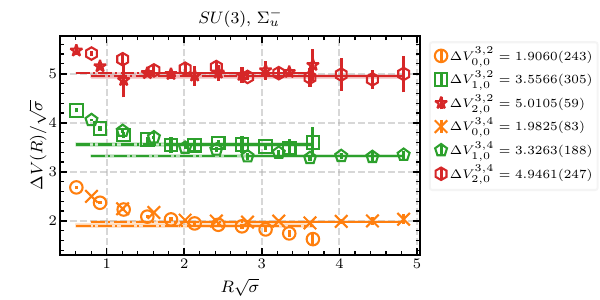}
	\includegraphics[scale=0.7500]{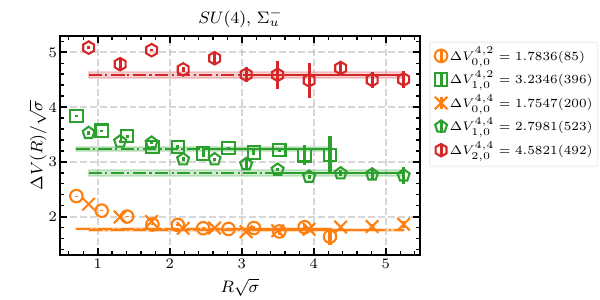}
	\includegraphics[scale=0.7500]{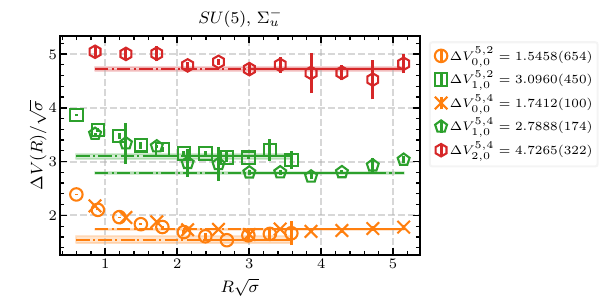}
	\includegraphics[scale=0.7500]{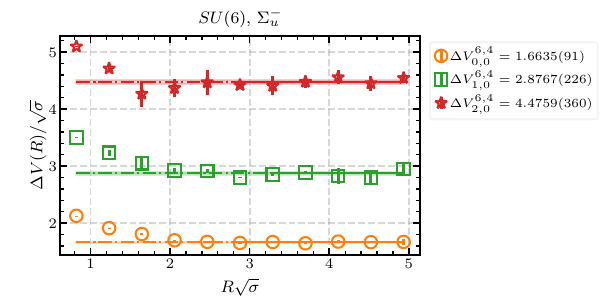}
\caption{We subtract the groundstate potential to find evidence for axions, for different groups. $\Delta V^{m, n}_{p, 0}$ indicates the data from ensemble $W_{m, n}$ in Table \ref{tab:ensemble}, and $V_{{\Sigma_u^-}^p}-V_{\Sigma_g^+}$ where $p$ is the excitation number.
\label{fig:axionplateau}}
\vspace{-0.55cm}
\end{figure}

As we examine higher excited states, we observe that some begin to show significant deviations from Nambu-Goto behavior, with stricking pattern seen in the ground state of $\Sigma_g^-$ and across the entire spectrum of states for $\Sigma_u^-$. These states exhibit behavior akin to that of a ground state with an additional constant mass term coupled, as illustrated in Figs. \ref{fig:SU5spectra} and \ref{fig:SU6spectra}. This behavior aligns with expectations for an axion coupled to the string. In Fig.~\ref{fig:axionplateau}, we highlight the presence of axion states in these open flux tubes, with the lightest possessing mass same as the axion observed in the closed flux tube spectra.

\begin{comment}
\begin{figure}[t!]
	\includegraphics[scale=0.500]{topological_charge_osc_A_0_B24.0_X4.0.pdf}
\caption{Topological charge history for $SU(6)$.\label{fig:tc_oscillation}}
\end{figure}
\end{comment}

%SSSSSSSSSSSSSSSSSSSSSSSSSSSSSSSSSSSSSSSSSSSSSSSSSSSSSSSSSSSSSSSSSSSSSSSSSSSSSSSSSSSSSSSSSSSSSSSSSSSSSSSSSSSSSS
%SSSSSSSSSSSSSSSSSSSSSSSSSSSSSSSSSSSSSSSSSSSSSSSSSSSSSSSSSSSSSSSSSSSSSSSSSSSSSSSSSSSSSSSSSSSSSSSSSSSSSSSSSSSSSS
%SSSSSSSSSSSSSSSSSSSSSSSSSSSSSSSSSSSSSSSSSSSSSSSSSSSSSSSSSSSSSSSSSSSSSSSSSSSSSSSSSSSSSSSSSSSSSSSSSSSSSSSSSSSSSS
\section{Conclusion}

In this study, we conducted a comprehensive analysis of the spectrum of the open flux tube across various discrete symmetry configurations and a large range of excitation levels. Specifically, we extracted the spectrum for $SU(3)$ gauge theories at two lattice spacings, $SU(4)$ and $SU(5)$ each at two lattice spacings, and $SU(6)$ at one lattice spacing. Our results show minimal finite lattice spacing effects and $N_c$ effects, indicating that the observed physics closely represents the large-$N_c$ limit and the continuum.

Our findings reveal that many states in the low-lying spectrum of the flux tube can be well approximated by the Nambu-Goto string model with moderate corrections. However, several states exhibit significant deviations from the Nambu-Goto string behavior, displaying characteristics of massive excitations. Notably, a number of these massive "axionic" excitations were observed, with the lightest having a mass matching its counterpart found in the close flux tube.

\vspace{-0.5cm}
%SSSSSSSSSSSSSSSSSSSSSSSSSSSSSSSSSSSSSSSSSSSSSSSSSSSSSSSSSSSSSSSSSSSSSSSSSSSSSSSSSSSSSSSSSSSSSSSSSSSSSSSSSSSSSS
%SSSSSSSSSSSSSSSSSSSSSSSSSSSSSSSSSSSSSSSSSSSSSSSSSSSSSSSSSSSSSSSSSSSSSSSSSSSSSSSSSSSSSSSSSSSSSSSSSSSSSSSSSSSSSS
%SSSSSSSSSSSSSSSSSSSSSSSSSSSSSSSSSSSSSSSSSSSSSSSSSSSSSSSSSSSSSSSSSSSSSSSSSSSSSSSSSSSSSSSSSSSSSSSSSSSSSSSSSSSSSS
\section*{Acknowledgments}
\vspace{-0.25cm}
	AS was supported by CEFEMA UIDB/04540/2020 Post-Doctoral Research Fellowship. AA acknowledges support by the ``EuroCC" project funded by the ``Deputy Ministry of Research, Innovation and Digital Policy and the Cyprus Research and Innovation Foundation" as well as by the EuroHPC JU under grant agreement No 101101903. The authors thank CeFEMA, an IST research unit whose activities are partially funded by FCT contract  UIDB/04540/2020 for R\&D Units. AA is also indebted to Conghuan Luo for performing a critical review of the manuscript.

    %\nocite{*} % Insert publications even if they are not cited in the poster
  %\renewcommand\refname{}% to remove References from the title of bibiliopgraphy

  \newpage
    \bibliography{references}
  \bibliographystyle{JHEP}

\end{document}